\documentclass{article}

\usepackage{graphicx}
\usepackage{amssymb}
\usepackage{amsmath}
\usepackage{mathptmx}
\usepackage{natbib}
\usepackage{color}
\usepackage{bm}

\title{Data-driven topology design using a deep generative model}

\author{Shintaro Yamasaki$^\text{a,}\footnote{Corresponding author: {\tt yamasaki@mech.eng.osaka-u.ac.jp} (Shintaro Yamasaki)}$ ,
        Kentaro Yaji$^\text{a}$,
        Kikuo Fujita$^\text{a}$ \\[12pt]
$^\text{a}$\textit{Department of Mechanical Engineering, Graduate School
              of Engineering,}\\
              \textit{Osaka University, 2-1 Yamadaoka,
              Suita 565-0871, Japan}}

\begin{document}

\maketitle

\begin{abstract}
In this paper, we propose a sensitivity-free and multi-objective structural design methodology called \textit{data-driven topology design}.
It is schemed to obtain high-performance material distributions from initially given material distributions in a given design domain.
Its basic idea is to iterate the following processes: (i) selecting material distributions from a dataset of material distributions according to eliteness, (ii) generating new material distributions using a deep generative model trained with the selected elite material distributions, and (iii) merging the generated material distributions with the dataset.
Because of the nature of a deep generative model, the generated material distributions are diverse and inherit features of the training data, that is, the elite material distributions.
Therefore, it is expected that some of the generated material distributions are superior to the current elite material distributions, and by merging the generated material distributions with the dataset, the performances of the newly selected elite material distributions are improved.
The performances are further improved by iterating the above processes.
The usefulness of data-driven topology design is demonstrated through numerical examples.
\flushleft

\end{abstract}

\section{Introduction}
\label{section_introduction}
Structural design is to determine the structural shape and topology of artifacts on the basis of physics, mathematics, designer intuition, and so on.
Among previously proposed methodologies for structural design, topology optimization originated by \citet{1988_bendsoe_01} is a promising one because of its potential to yield high-performance structures while considering both shape and topology.

There are two basic concepts in topology optimization \citep{2003_bendsoe_01}.
One is replacing a structural design problem with a material distribution problem in a given design domain.
The other is exploiting the optimal, or at least local optimal material distribution using sensitivity-based mathematical programming under a given objective function and constraints, that is, a given formulation.

Topology optimization has been applied to various engineering problems and has achieved immense success.
However, on the other hand, it includes some intrinsic difficulties.
Major one is the applicability to strongly nonlinear topology optimization problems.
That is, it is hard to directly solve some types of topology optimization problems because of their nonlinearity.
Representative problems are flow channel design problems under a turbulent flow \citep{2013_Kontoleontos_01,2018_dilgen_01}, and compliant mechanism design problems taking the geometric nonlinearity and the maximum stress into account \citep{2020_de_leon_01}.
Although a limited number of studies address solving these problems using sensitivity-based mathematical programming, it is also valuable to research alternative approaches.

Some sensitivity-free approaches may seem to be promising as alternative approaches.
For example, \citet{1994_chapman_01,2004_wang_03,2006_aguilar-madeira_01,2007_tai_02} proposed updating the structural shape and topology utilizing genetic algorithms (GAs).
Similarly, \citet{1997_shim_01} and \citet{2010_wu_01} proposed the use of simulated annealing and differential evolution, respectively.
Whereas they exploit solutions without the sensitivity exactly, their applicability is limited to relatively small-scale problems, as \citet{2011_sigmund_01} pointed out.
This limitation derives from the fact that it is difficult for sensitivity-free approaches to find optimal or satisfactory solutions within a realistic computational time, except for small-scale problems.

In other words, sensitivity-free approaches are useful if promising material distributions can be generated with a limited number of design variables.
On the basis of this point of view, \citet{2004_wang_03,2007_tai_02} introduced a graph-theoretic chromosome model to represent material distributions.
That is, in their approach, a parametric model representing material distributions is prepared in advance by considering the characteristics of the respective structural design problems.
However, introducing such a parametric model often means that the degree of freedom for representing material distributions is significantly restricted, which spoils the advantage of topology optimization.

Deep generative models~\citep{2013_kingma_01,2014_goodfellow_01} are promising to resolve this problem.
They are a type of generative models based on deep neural networks that include a low-dimensional space constructed with a small number of variables, called latent variables.
Features of training data are extracted through unsupervised learning, and similar but different data are generated by sampling in the aforementioned space called the latent space.
Because of the ability of deep neural networks, deep generative models can generate various material distributions with a large degree of freedom from a small number of latent variables.

Here, we consider training a deep generative model with diverse and promising material distributions.
The trained deep generative model generates material distributions that are diverse, while inheriting the features of the material distributions used as the training data.
Therefore, it is expected that some of them will be superior to the training data.
That is, a deep generative model has potential as an implicit parametric model for generating more promising material distributions by considering the latent variables as the design variables.

There are three important advantages of utilizing a deep generative model.
First, it is possible to conduct a solution search without the sensitivity information because of a small number of the latent variables, that is, the design variables.
Second, it is also possible to generate material distributions with a large degree of freedom, as described above.
Third, although it is expected that the relationship between the latent variables and the generated material distributions is very complex when satisfying both the above two points, this relationship is implicitly constructed through training.

As described above, our idea of utilizing a deep generative model relies on diverse and promising material distributions.
Here, we simply assume the multi-objective problem and regard material distributions whose non-dominated ranks are high (hereafter, we refer to them as high-rank material distributions) as diverse and promising material distributions.
That is, our idea discussed here is for multi-objective problems.

To obtain satisfactory solutions for multi-objective problems, we further introduce a data-driven approach.
That is, we scheme to update the training data by removing low-performance data entities and by adding high-performance data entities of the generated data.
We also scheme to iterate the data generation by a deep generative model and the update of the training data by the generated data.
When considering the training data as the elite data, the performances of the elite data will be gradually improved through the iteration, and the finally obtained elite data will be satisfactory solutions as a result of the performance improvement.

On the basis of the above discussions, in this paper, we propose to iteratively conduct the following processes after providing initial material distributions according to a certain policy: generating material distributions using a deep generative model trained with the current high-rank material distributions, merging the generated material distributions with the current high-rank material distributions, and newly selecting high-rank material distributions from the merged material distributions.
We call this sensitivity-free and multi-objective design methodology \textit{data-driven topology design}, because it aims to obtain satisfactory solutions rather than the optimal solutions and is based on a data-driven approach.
We demonstrate the usefulness of data-driven topology design by solving strongly nonlinear problems with an implemented method.

The rest of this paper is organized as follows.
We briefly introduce related studies in Section~\ref{section_related_work} and describe the framework in Section~\ref{section_framework}.
Next, we detail its implementation in Section~\ref{section_implementation_details} and provide numerical examples in Section~\ref{section_numerical_examples}.
Finally, we provide some concluding remarks in Section~\ref{section_conclusion}.

\section{Related studies}
\label{section_related_work}

\subsection{Topology optimization based on deep learning}
\label{section_TO_based_on_DD}
Recently, deep learning has gained significant attention from researchers in various fields, and some studies incorporating it into topology optimization have been proposed.
\citet{2016_ulu_01} proposed to predict optimized material distributions of the minimum compliance problem using a neural network.
In their study, various optimized material distributions were prepared using topology optimization while changing the load boundary condition.
The network is then trained under the load boundary condition as the input and the corresponding optimized material distribution as the output.
Using the trained network, the optimized material distribution for a given load boundary condition is predicted.

\citet{2019_zhang_01} also proposed to predict the optimized material distributions of the minimum compliance problem using a neural network.
In their study, the displacement and strain fields of the initial material distributions are used as the inputs, the corresponding optimized material distributions are used as the outputs, and the neural network is trained using the input and output data.
When an initial material distribution and its displacement and strain fields are given, the optimized material distribution is predicted using the trained network.
They demonstrated that their proposed method covers a change in the location where the displacement fixed boundary condition is imposed, in addition to the load boundary condition.

Some studies have focused on computational efficiency when predicting the optimized material distributions of the minimum compliance problem.
\citet{2019_cang_01} proposed to add training data entities on the basis of the Karush-Kuhn-Tucker conditions for improving the prediction accuracy while suppressing the computational cost.
\citet{2019_lei_01} proposed to utilize the framework of moving morphable components (MMC) \citep{2014_guo_01}.
Because MMC can represent structural shape and topology with a small number of design variables, it is expected that the computational cost for learning is reduced.
As a related study of \citet{2019_zhang_01}, \citet{2020_nie_01} proposed the utilization of stress and strain energy density fields of the initial material distributions as the input data.
In their study, a conditional generative adversarial network (cGAN) \citep{2014_mirza_01} is used to improve computational efficiency. 

Similar to the above studies, \citet{2019_yu_01} proposed a prediction method for the minimum compliance problem.
Optimized material distributions are predicted through two steps in their study.
First, an optimized material distribution under a given boundary condition is predicted in a low-resolution mesh, such as that described in the studies of \citet{2016_ulu_01,2019_zhang_01}.
Next, the predicted material distribution is refined in a high-resolution mesh using cGAN.

\citet{2020_abueidda_01} applied the concept of predicting optimized material distributions to nonlinear structural mechanics, in which nonlinear material is targeted.
Their proposed method predicts an optimized material distribution when the location, magnitude, and angle of the load are input.

As another notable application, \citet{2020_tan_01} proposed a prediction method for the design of microstructural materials.
Their method predicts microscale structures corresponding to the input material properties.
The design target of \citet{2019_zhang_02} is mask patterns in the photolithography process.
They trained the relationship between the mask patterns and the processed patterns using a variational autoencoder (VAE) \citep{2013_kingma_01}, and predicted the mask patterns corresponding to the desired patterns after the photolithography process.

The design target of \citet{2019_sasaki_01} is rotor structures of inner permanent magnet motors (IPMs).
In their study, quasi-optimal material distributions are exploited using a GA.
Although topology optimization incorporating a GA is generally time-consuming, it reduces the computational costs by utilizing a neural network that predicts the performance of IPMs.

Whereas many prediction methods for optimized material distributions focus on two-dimensional structural design problems, \citet{2018_banga_01} proposed a prediction method for three-dimensional problems.

Although these studies utilized deep learning for regression, some studies have focused on deep generative models.
\citet{2019_oh_01} proposed a topology optimization method for a wheel design problem in which the diversity of the optimized material distributions is ensured by referring to the material distributions generated by a generative adversarial network (GAN) \citep{2014_goodfellow_01}.
They also used an autoencoder \citep{2006_hinton_01} to evaluate the novelty of the optimized material distributions.

\citet{2018_guo_01} proposed a structural design method for the thermal compliance minimization problem, which consists of two steps.
First, a VAE is trained using various material distributions, which are obtained using topology optimization while changing the boundary conditions.
Next, the latent space of the trained VAE is exploited using a GA, and as a result of the exploitation, quasi-optimal material distributions are obtained.
In addition, a style transfer network \citep{2016_gatys_01} was used to reduce the noise included in the material distributions generated by the VAE.

\citet{2019_zhang_03} proposed a structural design method for the three-dimensional shape of a glider.
In their study, a VAE is trained using airplane models registered in a three-dimensional structure database \citep{2015_wu_01}, and the latent space of the trained VAE is exploited using a GA in a manner similar to that of \citet{2018_guo_01}.

Data-driven topology design may seem to be similar to the above studies, particularly the studies of \citet{2019_oh_01}, \citet{2018_guo_01}, and \citet{2019_zhang_03}.
However, its novelty can be clearly explained using estimation of distribution algorithm (EDA) \citep{2001_larranaga_01}.
Therefore, we introduce the EDA in the next section.

\subsection{Estimation of distribution algorithm}
\label{section_EDA}
Because of the generative nature for structures, data-driven topology design may seem to be an image-based GA in which only elite individuals are selected.
Indeed, this can be regarded as an EDA, which is a type of GA, on the basis of the following two points: (i) probabilistic models are constructed with elite individuals, and new individuals are generated using these models, and (ii) this generative process is iteratively performed.
Recently, \citet{2018_garciarena_01,2019_bhattacharjee_01} proposed the adoption of a VAE as a probabilistic model of an EDA, although their targets are well-studied test problems in the field of the GA rather than structural design problems.
The EDAs incorporating a VAE work well in their studies, which reinforces the validity of data-driven topology design.

Whereas the initial individuals are randomly generated in numerous studies on EDAs, data-driven topology design requires a policy for providing initial material distributions, which is explained in detail in Section~\ref{section_basic_concept}.
This is an important distinction between many studies conducted on EDAs and data-driven topology design.
Because the latter deals with material distributions represented with a large degree of freedom (typically, several thousands or more), it is difficult to prepare suitable initial material distributions using a random number generator.

\subsection{Novelty of data-driven topology design}
As discussed in Section~\ref{section_EDA}, data-driven topology design is novel in terms of its application to structural design and the policy for providing the initial individuals, when compared to previously proposed EDAs incorporating a deep generative model \citep{2018_garciarena_01,2019_bhattacharjee_01}.

Furthermore, data-driven topology design can be clearly distinguished from the studies of \citet{2019_oh_01}, \citet{2018_guo_01}, and \citet{2019_zhang_03}, from the viewpoint of an EDA.
That is, the former can be regarded as a type of EDA, whereas the latter cannot.
This is because a deep generative model is trained only by high-rank material distributions in the former, whereas various material distributions are used for training in the latter.
This is a critically important difference for our purpose, and we investigate the results caused by such a difference in Section~\ref{section_numerical_examples}.

\section{Framework}
\label{section_framework}

\begin{figure}[!t]
\begin{center}
\includegraphics[scale=1.0]{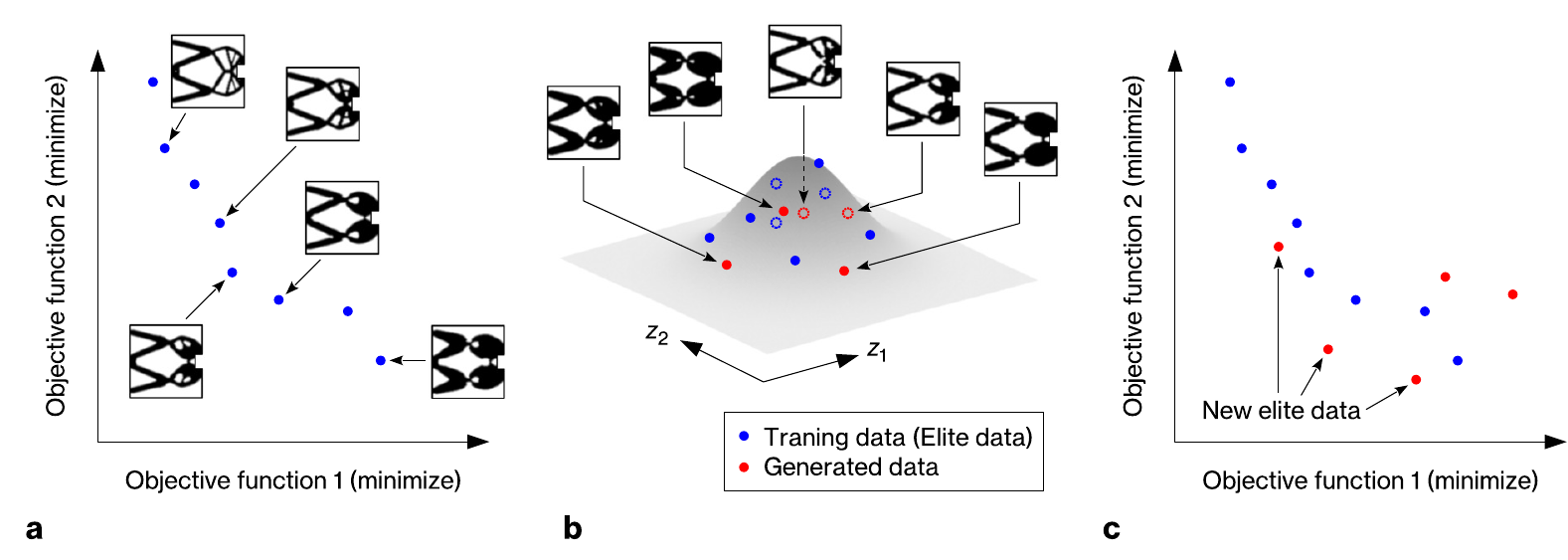}
\caption{Basic concept of data-driven topology design: \textbf{a} current elite data (high-rank material distributions) in multi-objective function space, \textbf{b} latent space constructed by training with elite data in \textbf{a} and generated data by sampling in this space, and \textbf{c} new elite data selected from generated data in \textbf{b}.
The height direction of \textbf{b} indicates the probability distribution of data, blue and red dotted circles indicate being on the invisible side, and $z_1$ and $z_2$ are latent variables
}
\label{fig_concept}
\end{center}
\end{figure}

\subsection{Basic concept}
\label{section_basic_concept}
In this section, we explain the basic concept of data-driven topology design in detail.
We illustrate it in Fig.~\ref{fig_concept}.
As described in Section~\ref{section_introduction}, data-driven topology design aims to generate higher-performance material distributions from already known high-rank material distributions in a multi-objective function space.
Figure~\ref{fig_concept}a shows an example of the already known high-rank material distributions, that is, the current elite data, in a structural design problem.
The elite data are used to train a deep generative model, and as a result, a latent space is constructed.
Figure~\ref{fig_concept}b shows an image of the latent space.
The elite data are located according to the probability distribution, and material distribution data are newly generated by sampling the latent space.
The generated data are evaluated in the multi-objective function space, and these with higher-performances are accepted as new elite data, as shown in Fig.~\ref{fig_concept}c.
Furthermore, old elite data dominated by the new elite data are removed from the elite data.
By doing so, we update the elite data.
These are further used as the training data to generate higher-performance material distributions.

As shown in Fig.~\ref{fig_concept}, the generated data are different from the training data, whereas they inherit the features of the training data.
It will be very difficult to construct an explicit parametric model with a small number of parameters, which can represent all of the training and generated data; however, we can construct such a parametric model implicitly by utilizing deep generative models.

\begin{figure}[!t]
\begin{center}
\includegraphics[scale=1.0]{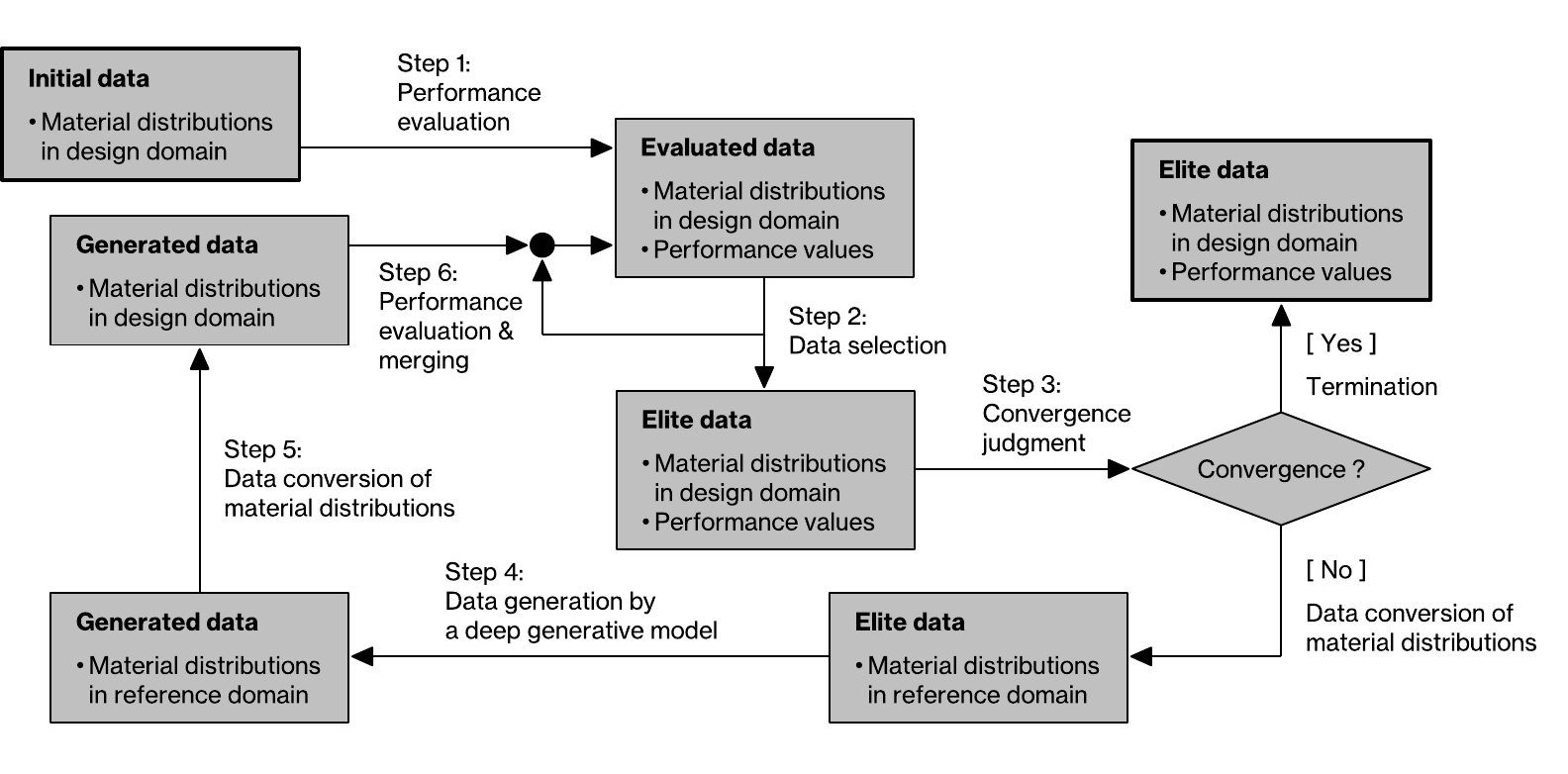}
\caption{Data process flow of data-driven topology design}
\label{fig_data_process_flow}
\end{center}
\end{figure}

On the other hand, data-driven topology design requires a policy for providing initial material distributions that are regular to a certain degree.
This is because deep generative models fail to capture meaningful features from randomly generated material distributions, which are usually extremely irregular.

To resolve this issue, we prepare material distributions by solving various topology optimization problems and select promising material distributions according to the optimality of the multiple objective functions.
Concretely, we propose two policies in this paper.

One is utilizing the outputs of a formulation support system, which was proposed by \citet{MP_2019_yamasaki_01} to support trial-and-error for determining an appropriate formulation of a topology optimization problem.
The formulation support system has a database constructed by collecting material distributions, which were obtained by solving various topology optimization problems.
A system user inputs multiple objective functions, and then the formulation support system outputs material distributions whose non-dominated rank is one (hereafter, we call them rank-one material distributions) by referring to the database.
Because material distributions in the database are results of topology optimization, it can be expected that these are regular to a certain degree.

The other is preparing material distributions by solving a topology optimization problem, which is easy to solve directly and is correlated with the target strongly nonlinear problem.
For example, if the target problem is a compliant mechanism design problem taking the geometric nonlinearity into account, we solve a compliant mechanism design problem assuming the linear strain while changing some parameters, such as the upper limits of constraints.
Then, we select promising material distributions from the viewpoint of optimality of the target problem.
In this policy, it is also expected that the selected material distributions are regular to a certain degree.

These policies stand on the viewpoint actively utilizing achievements in the field of topology optimization for solving strongly nonlinear problems.
In this paper, we use the former policy as the first choice, and if we fail to prepare sufficient initial material distributions by it, we use the latter policy.

\begin{figure}[!t]
\begin{center}
\includegraphics[scale=1.0]{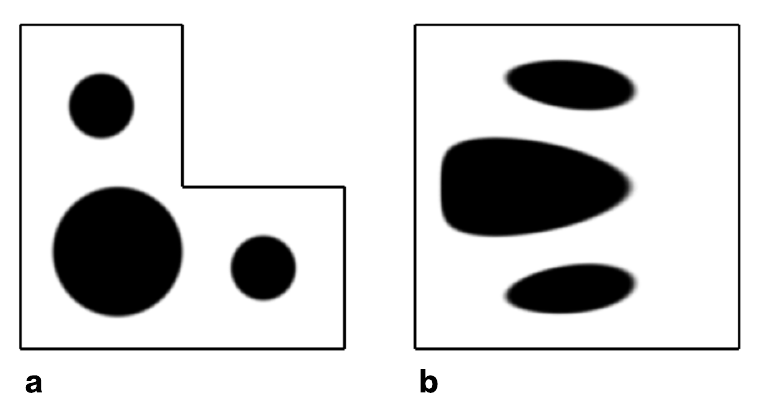}
\caption{Example of material distribution conversion using DDM: \textbf{a} material distribution in the design domain, and \textbf{b} converted material distribution conforming to the reference domain, where the material and void are shown in black and white, respectively}
\label{fig_DDM_ex}
\end{center}
\end{figure}

\subsection{Overall procedure}
\label{section_overall_procedure}
In this section, we describe the overall procedure of data-driven topology design.
This is schemed to obtain high-performance material distributions to a given design problem, which is defined by the shape of the design domain, boundary conditions, and multiple objective functions.
The data process flow starts from the preparation of the material distributions in the design domain, which are called \textit{the initial data}.
The initial data are provided according to a policy as described in Section~\ref{section_basic_concept}.

After preparing the initial data, these are processed as follows according to the indication in Fig.~\ref{fig_data_process_flow}.
\begin{description}
\item[\textbf{Step~1}] Evaluate the performances of the initial data by computing the values of the multiple objective functions.
Here, the data including the performance values are called \textit{the evaluated data}.
\item[\textbf{Step~2}] Select high-rank data entities from the evaluated data.
The selected data are called \textit{the elite data}.
Copies of the selected data are stored for merging with the generated data (see Step~6).
Note that, although we select only the rank-one data entities in the implementation of this paper, it is also possible to include other high-rank data entities.
\item[\textbf{Step~3}] Judge whether the elite data satisfy the convergence criterion (see details in Appendix~A).
If so, the current elite data are output as the final results.
Otherwise, the material distributions of the elite data are converted to conform to a normalized reference domain, which is a $1 \times 1$ square or $1 \times 1 \times 1$ cube in a two- or three-dimensional problem.
Such a conversion is applied because the normalized domain is suitable for image-based learning.
The design domain mapping (DDM) proposed by \citet{MP_2019_yamasaki_01} is used for the conversion.
Figure~\ref{fig_DDM_ex} shows an example of the material distribution conversion using the DDM.
\item[\textbf{Step~4}] Train a deep generative model using the converted material distributions and newly generate material distributions using the trained deep generative model.
These material distributions are called \textit{the generated data}.
\item[\textbf{Step~5}] Inversely convert the generated data to conform to the design domain, using the DDM.
\item[\textbf{Step~6}] Evaluate the performances of the generated data, in the same manner as step~1.
The generated data, including the performance values, are merged with the stored data of step~2.
The merged data are regarded as the evaluated data at the next iteration. Then, we return to step~2.
\end{description}
Through the above iterative procedure, we aim to obtain high-performance material distributions.

\section{Implementation details}
\label{section_implementation_details}
Although many deep generative models have recently been proposed, VAEs and GANs are representative.
When compared to a GAN, a VAE is suitable for data-driven topology design because its neural network architecture is relatively simple and a VAE is therefore robust \citep{2018_atienza_01}.
This robustness is particularly important because we train the neural network many times while updating the training data.
We therefore adopt a VAE as a deep generative model for the implementation.

Regarding the utilization of the VAE, some important implementation details are described in the following.

\subsection{Normalization of material distributions}
\label{section_normalization_of_material_distributions}
In data-driven topology design, we use two domains, that is, the design and reference domains, as described in Section~\ref{section_overall_procedure}.
In the design domain $D$, the material distributions are represented using the density function $\rho(\mathbf{x})$, where $\mathbf{x}$ are the coordinates of an arbitrary point in $D$.
$\rho(\mathbf{x})$ is continuous and takes a value of $0$ to $1$, and $\rho(\mathbf{x}) = 0$ and $1$ correspond to the void and material, respectively.
In contrast, $0 < \rho(\mathbf{x}) < 1$ corresponds to an intermediate state, according to the conventional manner of density-based topology optimization~\citep{1989_bendsoe_01}.
Similarly, the material distributions are represented using the density function $\rho(\boldsymbol{\xi})$ in the reference domain $\bar{D}$, where $\boldsymbol{\xi}$ are the coordinates of an arbitrary point in $\bar{D}$.

\begin{figure}[!t]
\begin{center}
\includegraphics[scale=1.0]{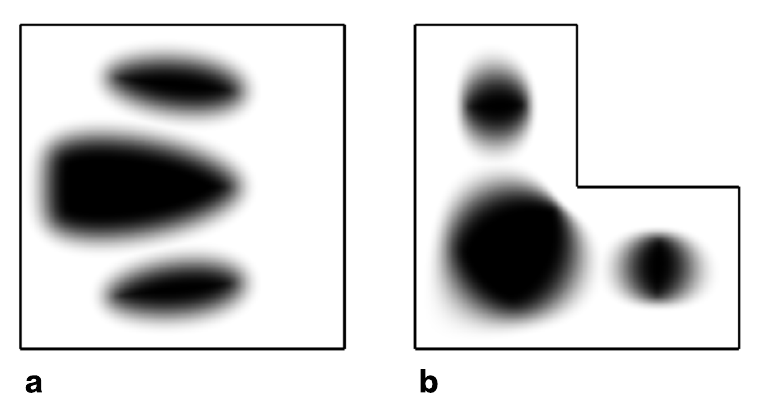}
\caption{Example of material distributions including wide transition zones: \textbf{a} material distribution normalized from that in Fig.~\ref{fig_DDM_ex}(b), and \textbf{b} material distribution in the design domain, which is inversely converted from that in \textbf{a}}
\label{fig_DDM_reg}
\end{center}
\end{figure}

When using the above representation model, we must consider the preferable features of the training data for the VAE.
In conventional density-based topology optimization, it is necessary to reduce the intermediate state while maintaining the smoothness of the material distribution.
From this perspective, the material distributions in Fig.~\ref{fig_DDM_ex}, for example, are preferable.
In contrast, it is thought that the intermediate state has a positive effect when training the VAE because it provides information regarding the outline of the structure.
In fact, MNIST~\citep{2012_deng_01}, one of the most important dataset in the field of deep learning, includes thousands of grayscale images of handwritten digits.

Therefore, we blur the outline in the reference domain $\bar{D}$ as follows.
First, we compute a scalar function $\phi(\boldsymbol{\xi})$ as
\begin{equation}
\phi(\boldsymbol{\xi}) = 2 \rho(\boldsymbol{\xi}) - 1.
\end{equation}
Next, we give $\phi(\boldsymbol{\xi})$ the signed distance characteristic to the iso-contour of $\phi(\boldsymbol{\xi}) = 0$, using a geometry-based re-initialization scheme \citep{MP_2010_yamasaki_03}.
Finally, we update $\rho(\boldsymbol{\xi})$ using the following equation:
\begin{equation}
\rho(\boldsymbol{\xi}) = \left\{
\begin{array}{cl}
 0 & \; \quad ( \; \phi(\boldsymbol{\xi}) < -h \; ), \\
 H(\phi(\boldsymbol{\xi})) & \; \quad ( \; -h \leq \phi(\boldsymbol{\xi}) \leq h \; ), \\
 1 & \; \quad ( \; h < \phi(\boldsymbol{\xi}) \; ), \\
\end{array} \right. \label{eq_heavd1}
\end{equation}
where $h$ is the parameter for the bandwidth of the transition zone from the void to the material, and $H(\phi)$, which is differentiable in $\left[-h, h \right]$ and $\frac{\mathrm{d}H}{\mathrm{d}\phi} = 0$ at $\phi = -h \; \mathrm{and} \; h$, is given as follows:
\begin{equation}
H(\phi) =
 \frac{1}{2}+\frac{15}{16}\Bigl(\frac{\phi}{h}\Bigr)-\frac{5}{8}\Bigl(\frac{\phi}{h}\Bigr)^{3} +\frac{3}{16}\Bigl(\frac{\phi}{h}\Bigr)^{5} \mathrm{.} \label{eq_heavd2}
\end{equation}
This process is a type of normalization to the material distribution; as an example, the material distribution in Fig.~\ref{fig_DDM_ex}b is processed, as shown in Fig.~\ref{fig_DDM_reg}a by setting $h$ to $0.08$.

Because of the normalization, the material distributions of the training data include wide transition zones from the void to the material (see Fig.~\ref{fig_DDM_reg}a).
Therefore, it is expected that the material distributions generated by the VAE also include wide transition zones.
If such material distributions are inversely converted into the design domain, the wide transition zones still remain, as shown in Fig.~\ref{fig_DDM_reg}b.
Because such wide transition zones often cause fatal numerical errors in the forward analysis, we need to binarize the material distributions in the design domain.
This is conducted by applying the normalization process described earlier to the design domain by setting $h$ to a small value.

We set $h$ to $0.08$ for the normalization in the reference domain on the basis of a preliminary study.
As demonstrated in Section~\ref{section_numerical_example_1}, the normalization method discussed here contributes to the generation of smooth material distributions.

\subsection{Details of data generation using VAE}
\label{section_details_VAE}
\begin{figure}[!t]
\begin{center}
\includegraphics[scale=1.0]{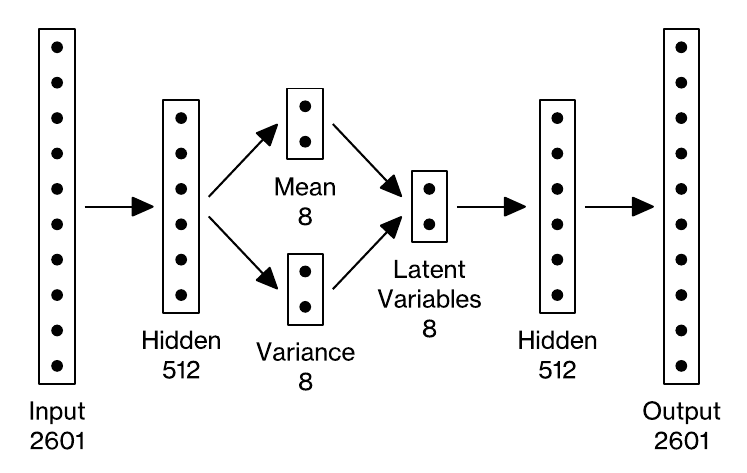}
\caption{Architecture of VAE}
\label{fig_VAE_arch}
\end{center}
\end{figure}

Figure~\ref{fig_VAE_arch} shows the architecture of the VAE used in the numerical examples of Section~\ref{section_numerical_examples}.
As shown in the figure, this is a type of multilayer perceptron that includes two hidden layers.
The reference domain $\bar{D}$ is discretized with $50 \times 50$ square elements, and the material distributions in $\bar{D}$ are represented using the values of the density function $\rho(\boldsymbol{\xi})$ at the lattice points.
Therefore, the input layer has $2,601$, that is, $51 \times 51$ neurons.
This input layer is fully connected to a hidden layer having $512$ neurons.
We referred the study of \citet{2018_atienza_01} to determine the size of the hidden layer.

After activating these neurons using the ReLU function, this layer is also fully connected to two layers having $8$ neurons, one corresponds to $\boldsymbol{\mu}$, which is the mean value vector of the latent variables $\mathbf{z}$, and the other corresponds to $\log{(\boldsymbol{\sigma}\circ\boldsymbol{\sigma})}$, where $\boldsymbol{\sigma}$ is the variance vector of $\mathbf{z}$, and $\circ$ represents the element-wise product.
We then obtain the latent variables $\mathbf{z}$ as follows:
\begin{equation}
\mathbf{z} = \boldsymbol{\mu} + \boldsymbol{\sigma}\circ\boldsymbol{\epsilon},
\end{equation}
where $\boldsymbol{\epsilon}$ is a random vector according to the standard normal distribution.
The number of the latent variables $N_{\mathrm{lt}}$ is an important parameter and we set it to 8 on the basis of a preliminary study.
The influence of this parameter to the obtained results is investigated in Appendix~B.

The layer of the latent variables $\mathbf{z}$ is further fully connected to a hidden layer having $512$ neurons.
After activating these neurons using the ReLU function, this layer is fully connected to the output layer having $2,601$ neurons, and outputs are obtained after sigmoid activation.
The outputs are interpreted as material distributions in $\bar{D}$ in the same manner as the inputs.
Note that the architecture described in this section is for two-dimensional problems.
If extending the architecture to three-dimensional problems, we may need additional dimensionality reduction techniques for larger inputs and outputs.

The VAE with the above architecture is trained using the elite data as the inputs and outputs, and the latent space composed of the latent variables is constructed through training.
In more detail, the training is conducted by minimizing the following loss function $L$ using the Adam optimizer \citep{2014_kingma_01}:
\begin{equation}
L := L_{\mathrm{recon}} + L_{\mathrm{KL}},
\label{eq_L}
\end{equation}
where $L_{\mathrm{recon}}$ is the reconstruction loss measured by the mean-squared error, and $L_{\mathrm{KL}}$ is a term corresponding to the Kullback-Leibler (KL) divergence.
$L_{\mathrm{KL}}$ is computed as follows:
\begin{equation}
L_{\mathrm{KL}} = -\frac{1}{2} \sum^{N_{\mathrm{lt}}}_{i=1} \left(1+\log \left( \sigma_{i}^2 \right) - \mu_{i}^2 -\sigma_{i}^2 \right),
\end{equation}
where $\mu_{i}$ and $\sigma_{i}$ are the $i$-th components of $\boldsymbol{\mu}$ and $\boldsymbol{\sigma}$, respectively.
The mini batch size and the learning rate are set to $20$ and $1 \times 10^{-3}$, respectively.

\begin{figure}[!t]
\begin{center}
\includegraphics[scale=1.0]{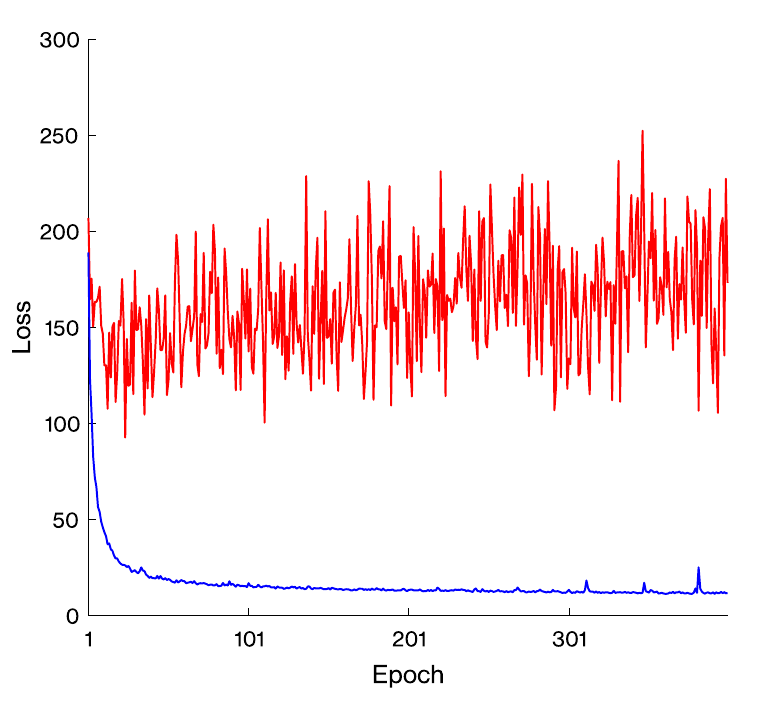}
\caption{Learning histories at iteration~0 of an example: loss function $L$ of the training data when all elite data are used as the training data (blue), and that of the validation data when half of the elite data are used as the validation data (red)}
\label{fig_loss_history}
\end{center}
\end{figure}

Because the dimensionality is drastically reduced from the input and output layers into a low-dimensional latent space, it is expected that important features of the training data are extracted into this space.
Furthermore, the range of the latent space that we should focus on is restrictive because the latent variables corresponding to the training data do not take extremely large or small values according to the probability distribution $N(0, 1)$.

On the basis of the above discussion, we generate material distributions by random sampling in the latent space.
The sampling vectors are composed of uniformly distributed random numbers in $\left[-4, 4 \right]$, and the material distributions are output from the sampling vectors using the trained VAE.
Thus, we generate 400 material distributions that are diverse and inherit the important features of the training data.

\subsection{Training and validation data for VAE}
In data-driven topology design, the VAE is iteratively trained while updating the training data.
Therefore, it is preferable to prepare a constant number of training data at every iteration.
By doing so, we can use the same parameter settings through the iterations for training the VAE.

For this reason, we provide $400$ material distributions as the training data at every iteration.
If the number of elite data entities is over $400$, we reselect $400$ data entities according to the crowding distance \citep{2002_deb_01} in the objective function space, as the elite data.
Otherwise, we make up for the shortage by making copies.

It should also be described that we do not prepare the validation data in the implementation discussed here.
In general, the validation data are used to avoid overfitting the training data.
On the other hand, VAE originally has a mechanism to avoid overfitting, as pointed out by \citet{WEB_2019_rocca_01}; that is, the regularization by introducing the KL divergence in (\ref{eq_L}).

In addition, the number of elite data entities is often extremely small (less than 100) in early iterations, and therefore, the number of training data entities further decreases if we prepare validation data.
In such a situation, it is almost meaningless to monitor the loss function $L$ of the validation data.
Figure~\ref{fig_loss_history} shows the histories of the loss function at iteration~0 of example~1, which is investigated in Section~\ref{section_numerical_example_1}.
The blue line indicates the case in which all elite data are used as the training data.
As shown in this figure, the loss function of the training data converges almost smoothly.
In contrast, when half of the elite data are used as the validation data, the loss function of the validation data violently vibrates, as indicated by the red line.
In this case, it is almost impossible to judge whether the VAE is appropriately trained or not.
Therefore, we simply train the VAE for 400 epochs without the validation data at every iteration.

\section{Numerical examples}
\label{section_numerical_examples}

In this section, we provide three numerical examples to demonstrate the usefulness of data-driven topology design.
In example~1, we solve a high-stiffness and light-weight structure design problem, namely, the well-studied minimum compliance problem, to investigate the basic potential of data-driven topology design.
In this example, the formulation support system is used to provide the initial material distributions.

In example~2, we solve a low-stress and light-weight structure design problem as a strongly nonlinear problem.
The formulation support system is used to provide the initial material distributions, similar to example~1.
Whereas the main purpose of data-driven topology design is to obtain higher-performance material distributions from the initially provided material distributions, it brings secondary but important utility to the formulation support system itself.
We also discuss this utility.

Finally, in example~3, we tackle a compliant mechanism design problem taking the geometric nonlinearity and the maximum stress into account, as a further strongly nonlinear problem.
It was difficult for the current version of the formulation support system to provide a sufficient number of initial material distributions that function as compliant mechanisms.
Therefore, in this example, a simple compliant mechanism design problem assuming the linear strain is solved to provide the initial material distributions.

In all numerical examples, we use the International System of Units, assume the plane stress condition, and set the value of $h$ in (\ref{eq_heavd1}) to $0.025$ to binarize the material distributions in the design domain.
All numerical examples are computed using a workstation, which has 384GB memory and 36 cores.
The CPU model is Intel Xeon Gold 6240.

\begin{figure}[!t]
\begin{center}
\includegraphics[scale=1.0]{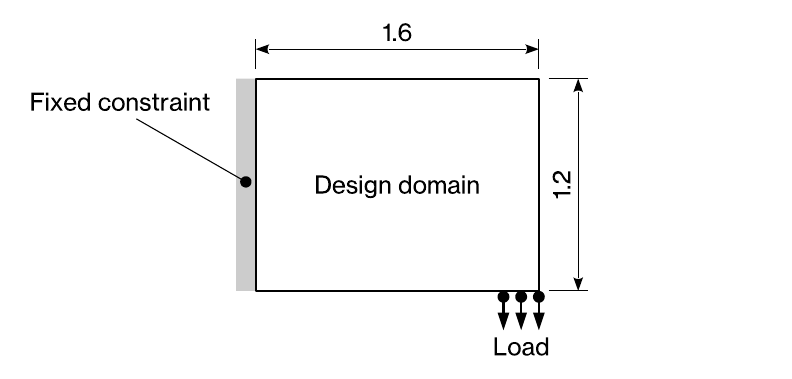}
\caption{Design domain and boundary conditions of example~1}
\label{fig_design_domains}
\end{center}
\end{figure}

\begin{figure}[!t]
\begin{center}
\includegraphics[scale=1.0]{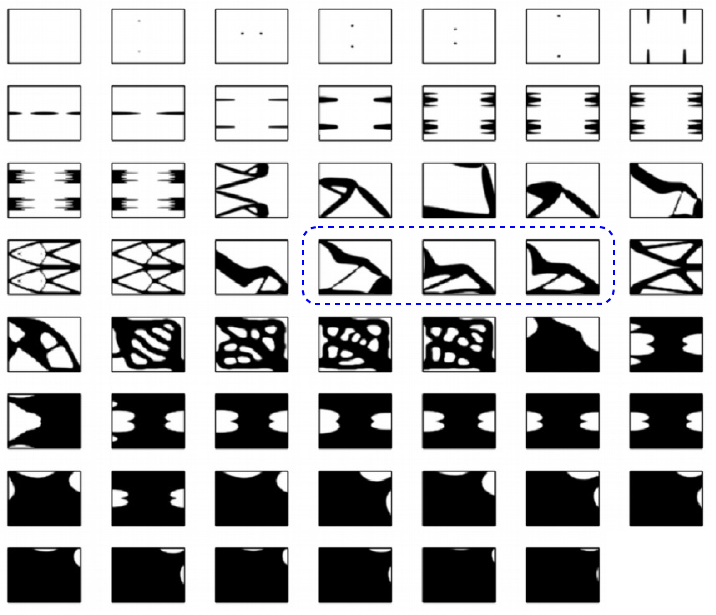}
\caption{Material distributions of the elite data at iteration~0 in example~1}
\label{fig_ex1_org_mat}
\end{center}
\end{figure}

\subsection{Example~1}
\label{section_numerical_example_1}

As described above, we solve the simple high-stiffness and light-weight structure design problem in example~1, the design domain and boundary conditions of which are shown in Fig.~\ref{fig_design_domains}.
As shown in this figure, a vertical load is applied to the bottom-right boundary and the displacement is fixed on the left-side boundary of the design domain.
The design domain is discretized with $128 \times 96$ square elements, and the magnitude of the applied load per unit area is set to $1$.
Young's modulus of the structural material is set to $1$ and is set to $1 \times 10^{-6}$ in the void to avoid the singular stiffness matrix, and Poisson's ratio is set to $0.3$.
In this example, two objective functions are set: one is the volume of the structure, and the other is the logarithm of the mean compliance to the applied load.

\begin{figure}[!t]
\begin{center}
\includegraphics[scale=1.0]{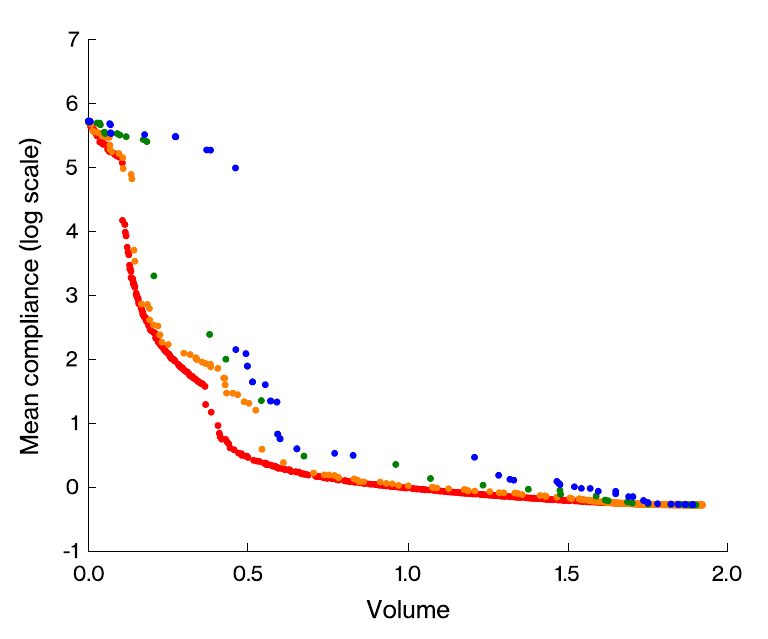}
\caption{Performances of elite solutions in example~1: iteration~0 (blue), iteration~1 (green), iteration~5 (orange), and iteration~38 (red)}
\label{fig_ex1_history}
\end{center}
\end{figure}

We construct the database of the formulation support system in a manner similar to that of \citet{MP_2019_yamasaki_01} and convert the material distributions in the database to conform to the design domain using the DDM.
By doing so, we provide $2,271$ material distributions as the initial data.
In step~1, we evaluate the volume and mean compliance of these material distributions using the finite element method (FEM) and obtain the evaluated data.
In step~2, we select the rank-one data entities from the evaluated data to obtain the elite data, and store their copies for the merging at step~6.
The material distributions of the elite data are shown in Fig.~\ref{fig_ex1_org_mat}.
In step~3, we convert these material distributions to conform to the reference domain using the DDM.
In step~4, material distributions are generated using the VAE, as described in Section~\ref{section_details_VAE}.
In step~5, we inversely convert the generated material distributions to conform to the design domain using the DDM.
In step~6, we evaluate the performances of the generated material distributions and merge them with the stored data and then return to step~2.
We iterate the above data generation procedure until the convergence criterion is satisfied at iteration~38.
The computational time per iteration is about 4 minutes.

Figure~\ref{fig_ex1_history} shows that the performances of the elite solutions gradually improve when iterating the data generation.
Because the performances are clearly improved after iteration~1, iterating the data generation procedure is significantly important for obtaining high-performance material distributions.

\begin{figure}[!t]
\begin{center}
\includegraphics[scale=1.0]{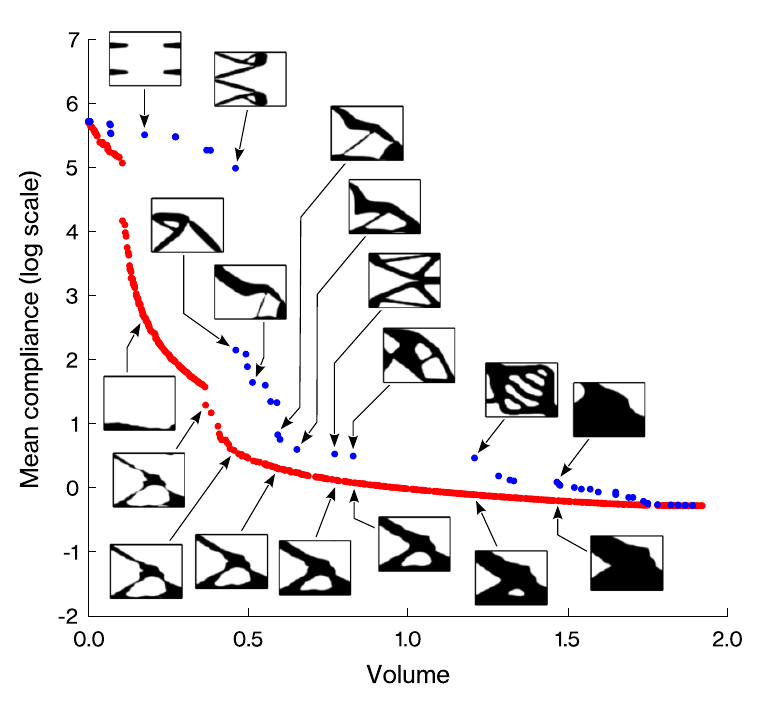}
\caption{Performances and representative material distributions of elite solutions at iteration~0 in example~1 (blue) and at iteration~38 (red)}
\label{fig_ex1_result}
\end{center}
\end{figure}

\begin{figure}[!t]
\begin{center}
\includegraphics[scale=1.0]{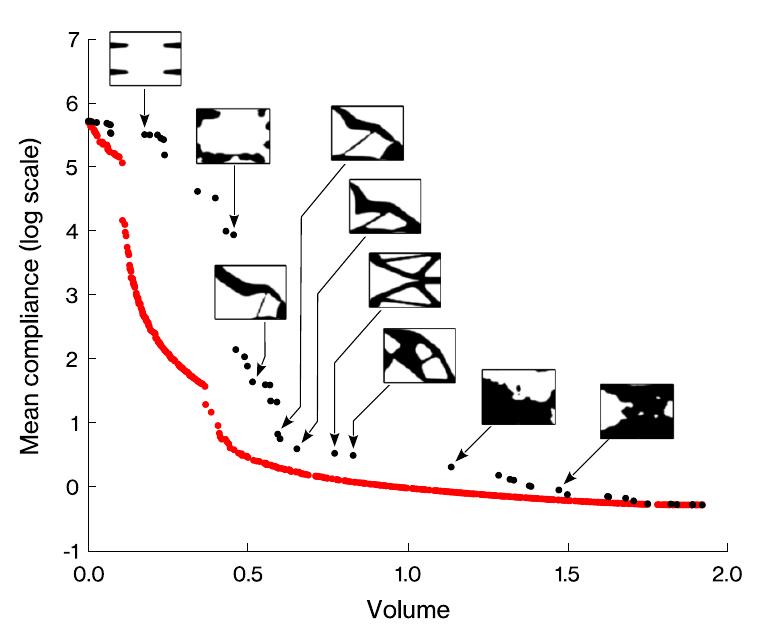}
\caption{Performances of elite solutions obtained by the proposed implementation in example~1 (red) and those obtained using a VAE trained with all of the material distributions (black), and representative material distributions of the latter}
\label{fig_ex1_extra1}
\end{center}
\end{figure}

Figure~\ref{fig_ex1_result} shows representative material distributions of the elite solutions at iterations~0 and 38.
As shown in this figure, the final material distributions obtained seem to be reasonable and similar to the well-known optimized structures of the minimum compliance problem, as discussed later, whereas the material distributions at iteration~0 are low-performance and therefore seem to be unreasonable.

Next, we discuss the importance of training the VAE using only the elite data.
For this purpose, we provide a case study where eliteness-based data selection is deactivated.
Figure~\ref{fig_ex1_extra1} shows the result of this case study.
Clearly, the performances of the obtained elite solutions are inferior to those obtained by the proposed implementation.
More importantly, the finally obtained elite material distributions seem to be very poor; in particular, some of them remain as elite material distributions from beginning to end.
These results indicate the disadvantage of training a VAE using all material distributions.
If a VAE is trained using all material distributions, various features of low-performance material distributions will be reflected in the latent space.
Therefore, it is extremely difficult to expect a VAE to efficiently generate high-performance material distributions with a limited number of sampling points.
Thus, the results shown in Fig.~\ref{fig_ex1_extra1} indicate the importance of training a VAE using only the elite data.

\begin{figure}[!t]
\begin{center}
\includegraphics[scale=1.0]{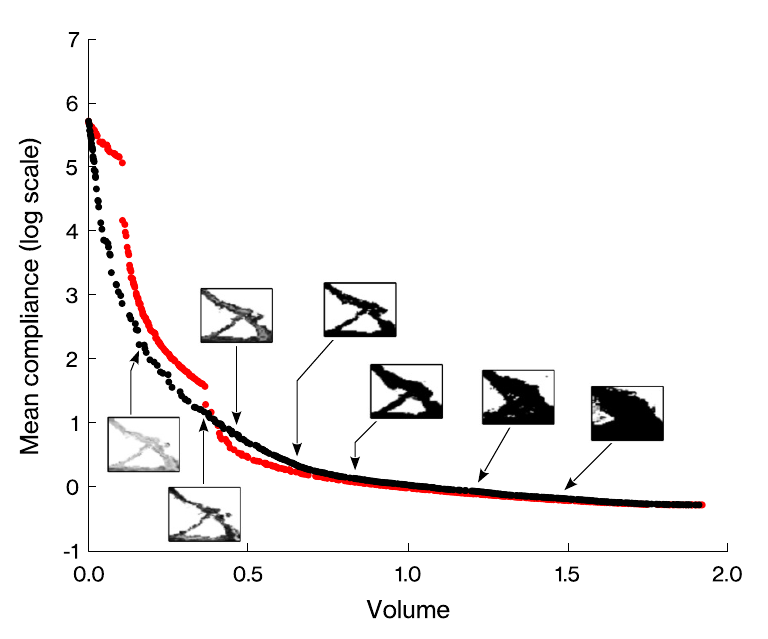}
\caption{Performances of elite solutions obtained by the proposed implementation in example~1 (red) and those obtained without the normalization method described in Section~\ref{section_normalization_of_material_distributions}(black), and representative material distributions of the latter}
\label{fig_ex1_extra3}
\end{center}
\end{figure}

\begin{figure}[!t]
\begin{center}
\includegraphics[scale=1.0]{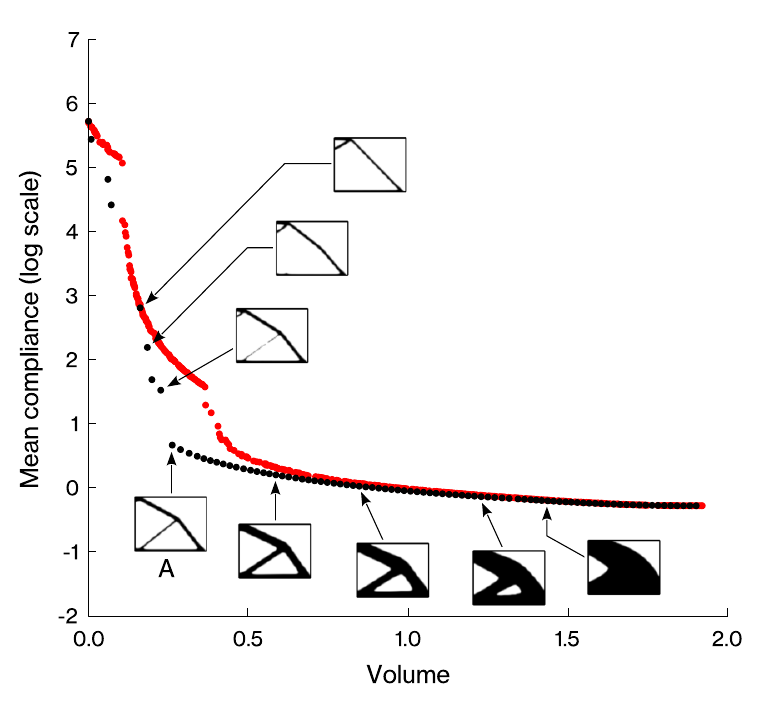}
\caption{Performances of elite solutions obtained by the proposed implementation in example~1 (red) and those obtained using density-based topology optimization (black), and representative material distributions of the latter}
\label{fig_ex1_extra2}
\end{center}
\end{figure}

Next, we investigate the usefulness of the normalization method introduced in Section~\ref{section_normalization_of_material_distributions}.
Figure~\ref{fig_ex1_extra3} shows the performances of the elite solutions obtained by the proposed implementation and those obtained without the normalization method.
As shown in this figure, the latter is superior to the former when the volume is less than $0.36$.
However, the material distributions of the latter are noisy images and include many intermediate states.
Because such characteristics are not preferable, we adopt the normalization method for data-driven topology design.

Finally, we compare the results obtained by the proposed implementation with those obtained by density-based topology optimization.
The elite solutions colored with black in Fig.~\ref{fig_ex1_extra2} are obtained by directly solving the well-known minimum compliance problem, while changing the allowable upper limit of the volume from $0.02$ to $1.90$ in increments of $0.02$.
These material distributions are normalized in the same manner as those obtained by the proposed implementation, and therefore, the performances of material distributions that include many elements in intermediate states are significantly deteriorated.
As a result, these material distributions are omitted from the elite solutions.
It is observed when the volume is less than $0.25$.

The solutions displayed with black in Fig.~\ref{fig_ex1_extra2} are close to the theoretically optimal solutions, and the elite solutions obtained by the proposed implementation are also close to them when the volume is greater than $0.5$ and less than $0.18$.
Furthermore, the representative material distributions shown in Figs.~\ref{fig_ex1_result} and \ref{fig_ex1_extra2} are also similar.
On the other hand, the proposed implementation fails to generate wire-like slender structures as material distribution A in Fig.~\ref{fig_ex1_extra2}.
Therefore, we consider that data-driven topology design has sufficient potential for generating high-performance material distributions although the above issue exists in the current implementation.
The material distributions encircled with the blue dotted line in Fig.~\ref{fig_ex1_org_mat} are slightly similar to the material distributions in Fig.~\ref{fig_ex1_extra2}, and we therefore assume that data-driven topology design updates elite solutions while propagating and improving these material distributions.

\subsection{Example~2}
\label{section_numerical_example_2}
\begin{figure}[!t]
\begin{center}
\includegraphics[scale=1.0]{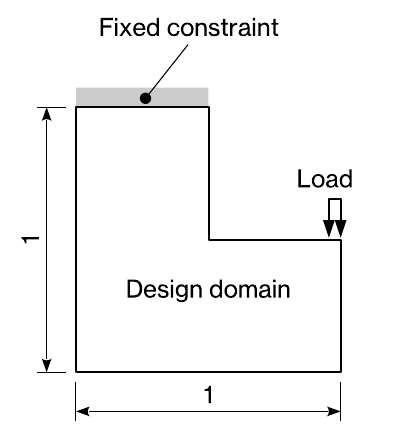}
\caption{Design domain and boundary conditions of example~2}
\label{fig_ex2_dd}
\end{center}
\end{figure}

In this section, we solve a low-stress and light-weight structure design problem, the design domain and boundary conditions of which are given in Fig.~\ref{fig_ex2_dd}.
As shown in this figure, a vertical load is applied to the center-right boundary and the displacement is fixed on the top boundary of the design domain.

For this design problem, two objective functions are set: the volume of the structure and the logarithm of the maximum value of the von Mises stress generated in the structure.
Furthermore, the mean compliance is imposed as a constraint to ensure the mechanical connection from the displacement fixed boundary to the load imposed boundary, which is crucial to obtain meaningful structures, as discussed by \citet{MP_2019_yamasaki_01}.

Although some studies have focused on the stress constraint or minimization problem, almost all of these studies introduce some type of approximation to avoid the point-wise constraints or min-max problem.
For example, \citet{2008_allaire_01,2013_holmberg_01,2015_de_leon_01} used global p-norm based stress measures, and \citet{2010_amstutz_01} also proposed a domain integral-type approximation.
These techniques are approximations, and it is often difficult to appropriately determine the artificial parameters used in them.
Furthermore, these studies allow intermediate states of material existence, although the stress on an intermediate state is often physically meaningless, as \citet{2013_holmberg_01} pointed out.
Because the stress usually takes the maximum value on the structural surface, intermediate states are originally unsuitable for the stress constraint or minimization problem.
Thus, the design problem discussed in this section is a difficult and strongly nonlinear problem because of the pointwise nature of the stress, when solving it in a strict manner.

\begin{figure}[!t]
\begin{center}
\includegraphics[scale=1.0]{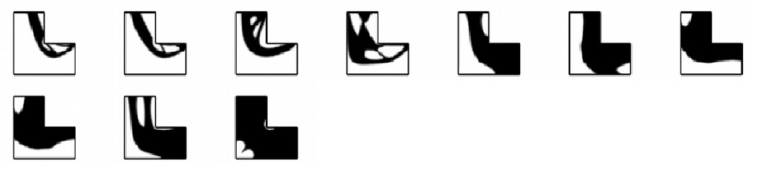}
\caption{Material distributions of the elite data at iteration~0 in example~2}
\label{fig_ex2_org_mat}
\end{center}
\end{figure}

The design domain is discretized with $78282$ triangular elements with the representative length of $0.005$, and the magnitude of the applied load per unit area is set to $1$.
The material properties are set to the same values as in example~1.
In this example, we use conforming meshes to structural boundaries proposed by \citet{MP_2017_yamasaki_03} to accurately compute the von Mises stress.

\begin{figure}[!t]
\begin{center}
\includegraphics[scale=1.0]{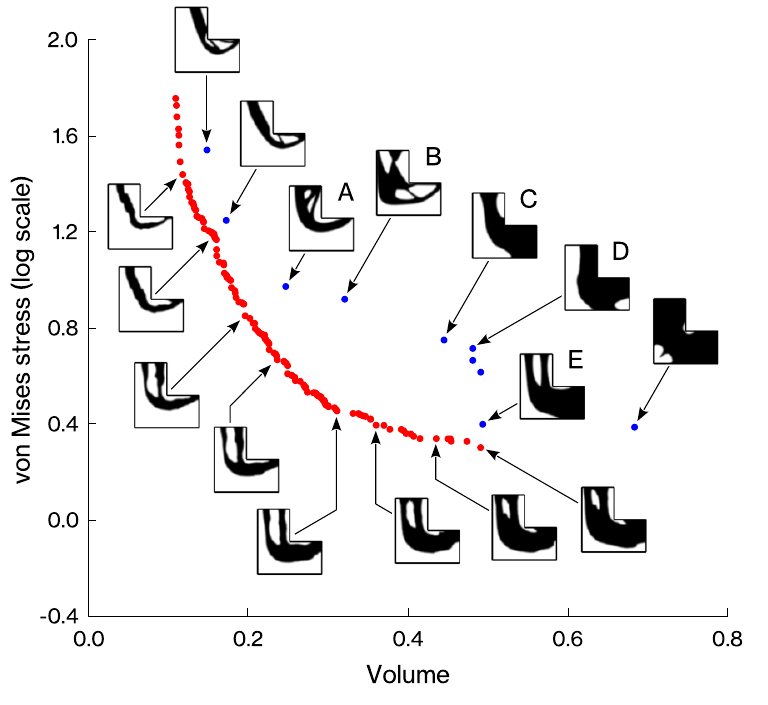}
\caption{Performances and representative material distributions of elite solutions at iteration~0 in example~2 (blue) and those at iteration~50 (red)}
\label{fig_ex2_result}
\end{center}
\end{figure}

We prepare the initial data in the same manner as in example~1 and compute the performances of the material distributions of the initial data, that is, the volume, maximum value of the von Mises stress, and mean compliance.
Next, we select the rank-one material distributions regarding the volume and maximum value of the von Mises stress under a constraint in which the mean compliance is less than 10.
Figure~\ref{fig_ex2_org_mat} shows the material distributions selected as the elite data.
In the same manner as in example~1, we iterate the data generation procedure until the convergence criterion is satisfied at iteration~50, and obtain the result shown in Fig.~\ref{fig_ex2_result}.
The computational time per iteration is about 7 minutes.

As shown in this figure, the performances of the elite solutions are drastically improved by data-driven topology design.
Furthermore, the obtained material distributions are reasonable, whereas some unreasonable material distributions exist at iteration~0.
For example, two holes of material distribution~A seem to be useless for avoiding stress concentration.
Material distribution~B also seems to be unreasonable because the narrow part on the top side of the design domain is not suitable to avoid stress concentration.
In material distributions~C and E, the material at the bottom-side of the design domain is not needed to support the load.
Furthermore, material distributions~C and D include an obvious singular point of the stress.

These unreasonable material distributions are suppressed as a result of the performance improvement of the elite solutions, and we can find some design knowledge from the results of data-driven topology design.
For example, structures such as the mirror writing of the letter "J" are preferable for avoiding stress concentration at the inner corner.
Furthermore, it is effective to reduce the material volume on the top-side of the design domain for weight saving.

The above results indicate that data-driven topology design provides an important utility to the formulation support system.
As described in Section~\ref{section_basic_concept}, the formulation support system is used to support the designer's trial-and-error for determining appropriate formulations of topology optimization problems.
That is, the designer sets multiple objective functions as the candidates of the objective and constraint functions in a topology optimization problem (the formulation candidate), and then, the formulation support system outputs the rank-one material distributions from the database as the blue solutions in Fig.~\ref{fig_ex2_result}.
By referring to these material distributions, the designer judges whether the formulation candidate is appropriate or not.

In the above situation, if useful design knowledge is obtained from reasonable material distributions as a result of the performance improvement by data-driven topology design, the designer can judge the appropriateness of the input formulation candidate with confidence.
Otherwise, the appropriateness of the input formulation candidate should be doubted.
In addition, design knowledge itself is useful for designers.
Therefore, we consider that data-driven topology design has potential, not only to obtain higher performance material distributions, but also to enhance the formulation support system.

\begin{figure}[!t]
\begin{center}
\includegraphics[scale=1.0]{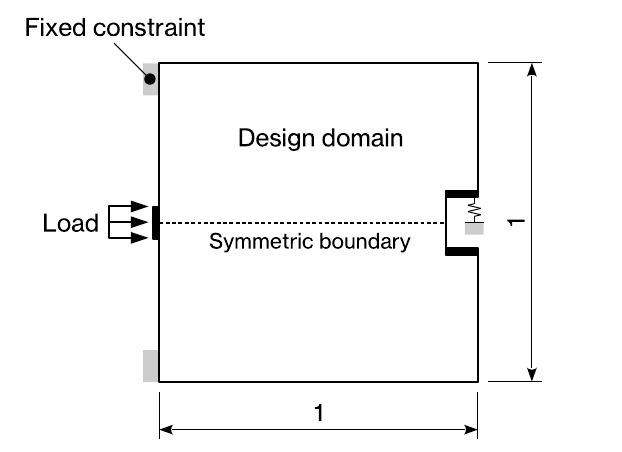}
\caption{Design domain and boundary conditions of example~3}
\label{fig_ex4_dd}
\end{center}
\end{figure}

\begin{figure}[!t]
\begin{center}
\includegraphics[scale=1.0]{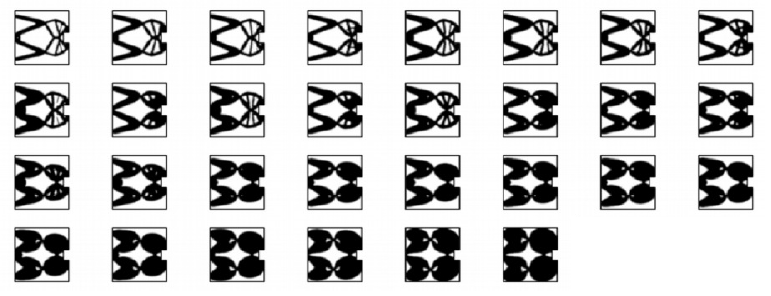}
\caption{Material distributions of the elite data at iteration~0 in example~3}
\label{fig_ex4_org_mat}
\end{center}
\end{figure}

\begin{figure}[!t]
\begin{center}
\includegraphics[scale=1.0]{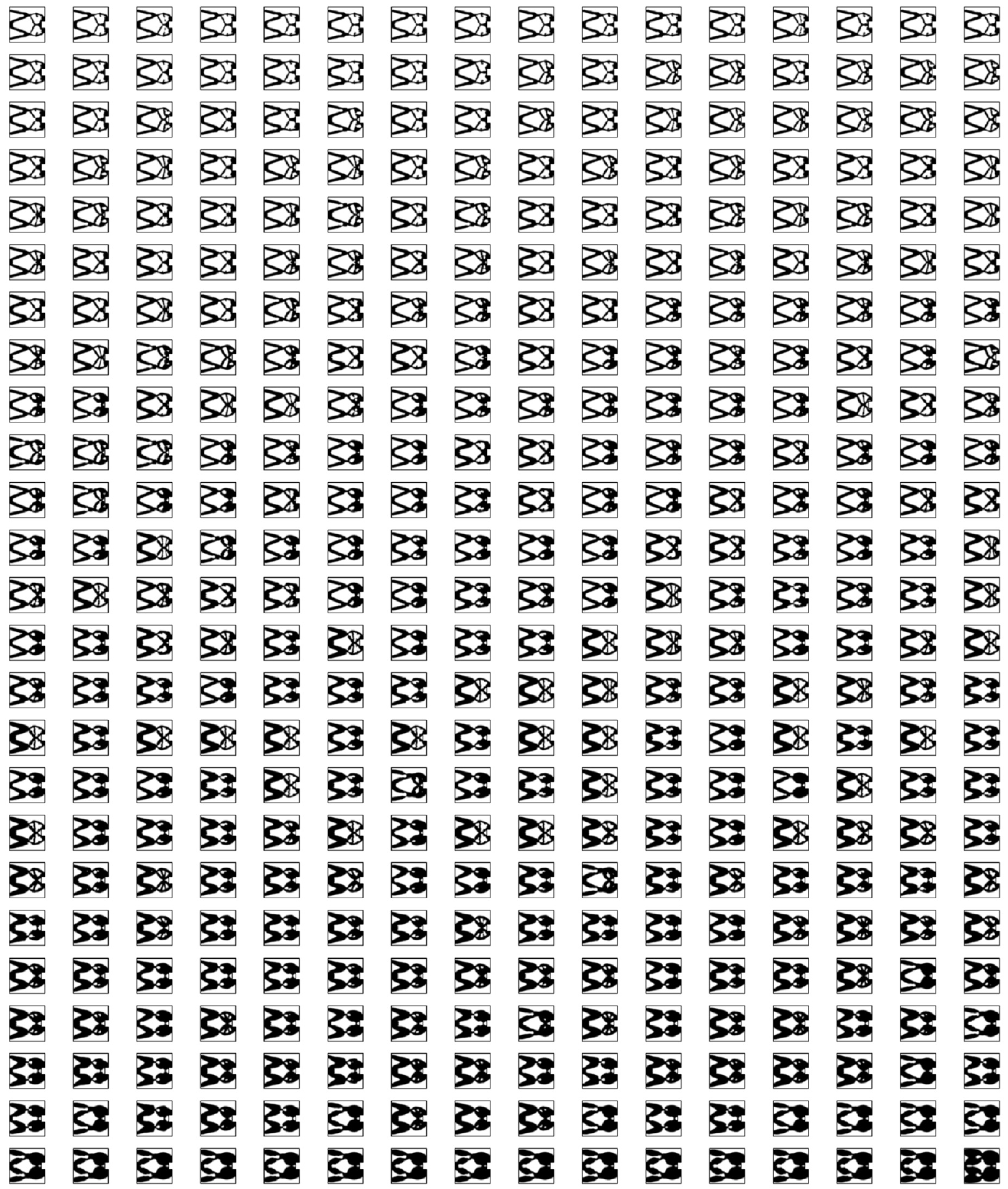}
\caption{Material distributions of the elite data at iteration~50 in example~3}
\label{fig_ex4_opt_mat}
\end{center}
\end{figure}

\begin{figure}[!t]
\begin{center}
\includegraphics[scale=1.0]{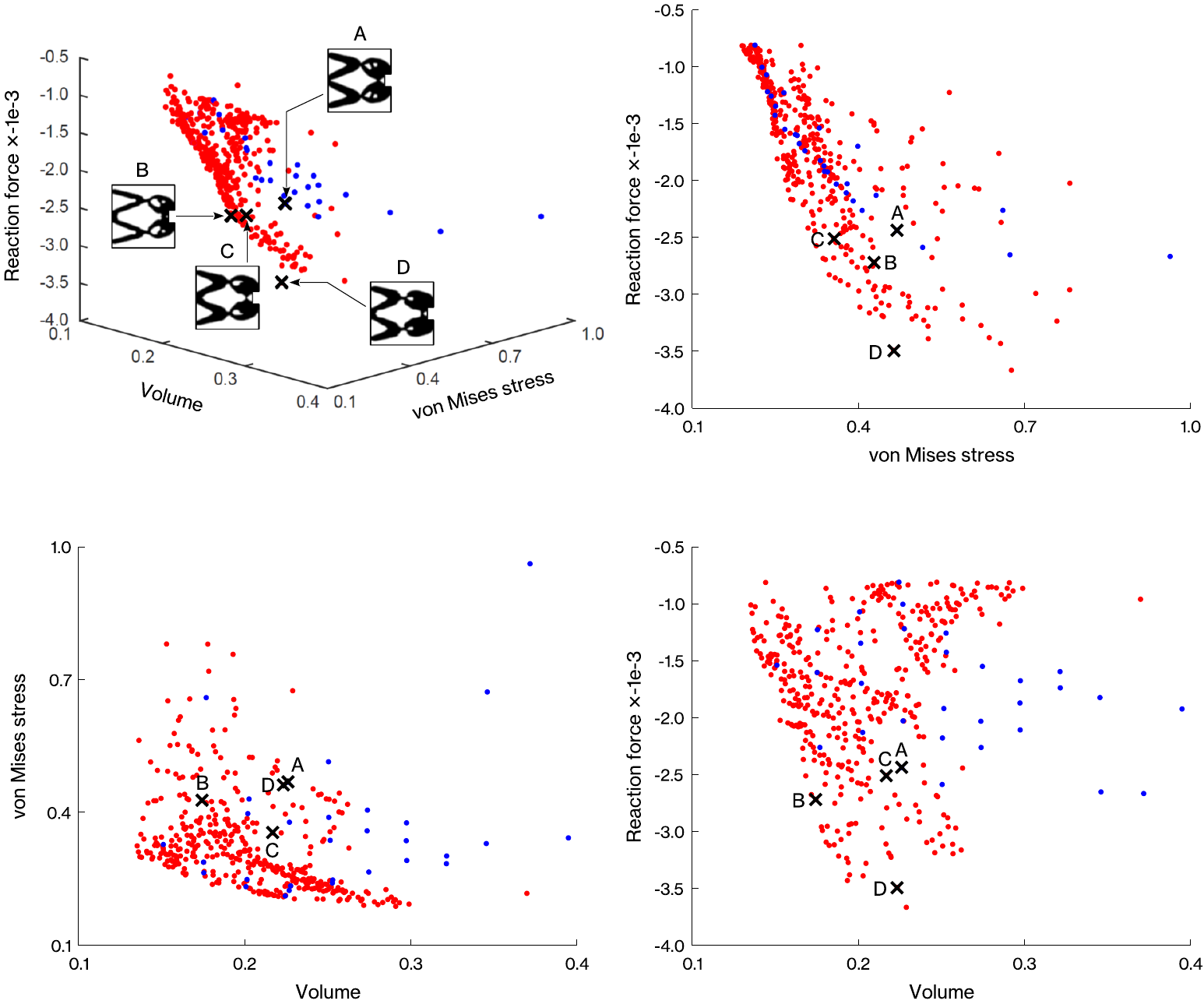}
\caption{Performances and representative material distributions of elite solutions at iteration~0 in example~3 (blue) and at iteration~50 (red)}
\label{fig_ex4_result}
\end{center}
\end{figure}

\begin{figure*}[!t]
\begin{center}
\includegraphics[scale=1.0]{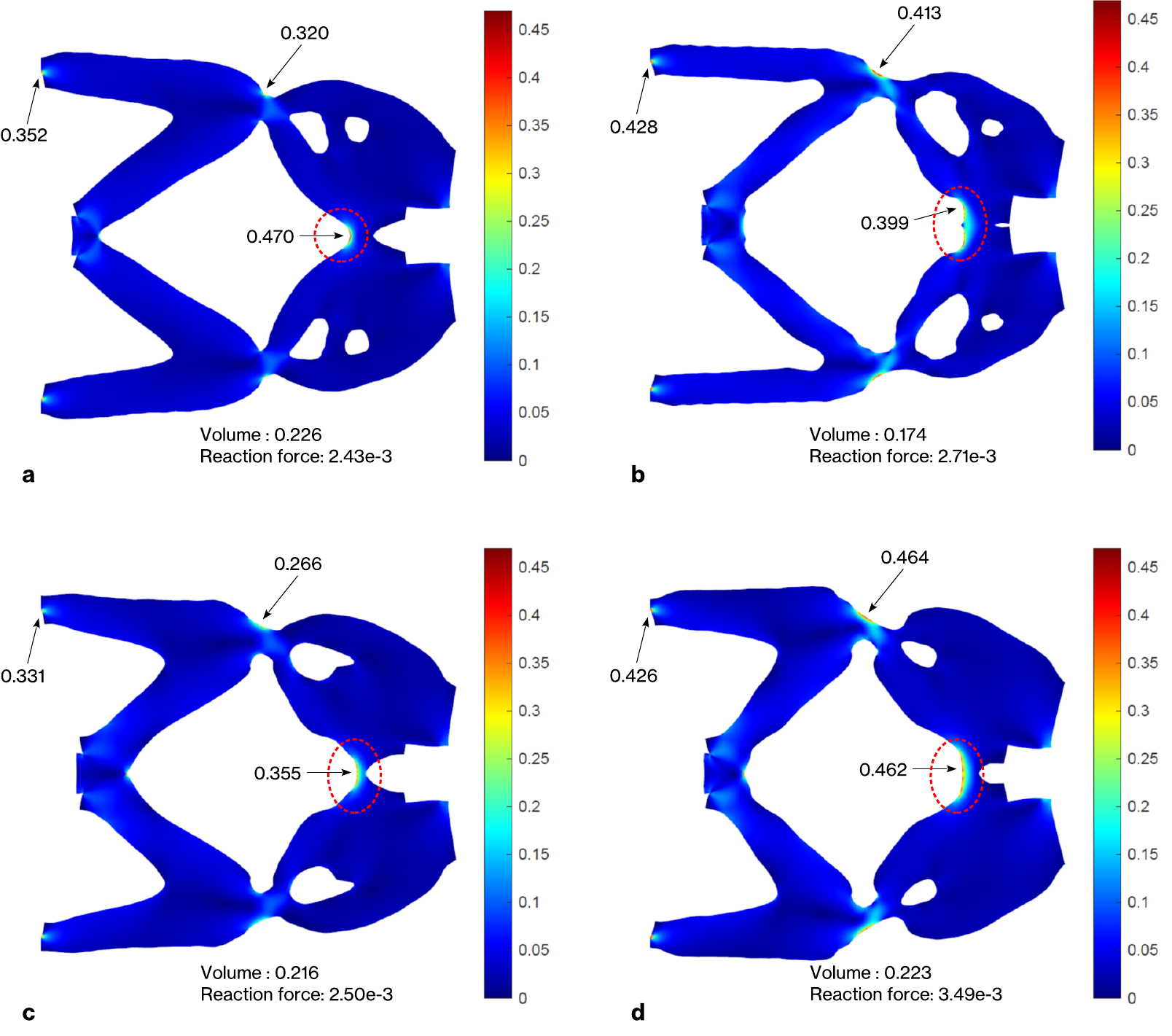}
\caption{Contour plots of von Mises stress with exact deformations for material distributions shown in Fig.~\ref{fig_ex4_result}: \textbf{a} -- \textbf{d} material distributions~A -- D}
\label{fig_ex4_detail}
\end{center}
\end{figure*}

\subsection{Example~3}
In example~3, we tackle a compliant mechanism design problem in which both the geometric nonlinearity and the maximum stress are considered.
Compliant mechanisms are well-studied design targets in the field of topology optimization; however, only a limited number of studies consider the geometric nonlinearity~\citep{2008_luo_03,2017_liu_01,2019_chen_02,2020_dunning_01,2020_kumar_01}, despite the large deformation of compliant mechanisms.
Similarly, only a limited number of studies consider the maximum stress as a constraint~\citep{2015_de_leon_01,2016_lopes_01}, although it is necessary to constrain the maximum stress in the real world.
Instead of constraining the maximum stress, many studies avoid the stress concentration in an indirect manner, for example, setting an artificial spring in the input port, constraining the stiffness to the input load, and so on.

The study of \citet{2020_de_leon_01} is the most challenging because the proposed method considers both the geometric nonlinearity and the maximum stress.
However, in their method, the maximum stress is approximately evaluated using the global p-norm, similar to \citet{2008_allaire_01,2013_holmberg_01,2015_de_leon_01}.

The above indicates that the compliant mechanism design problem that considers both the geometric nonlinearity and the maximum stress is a strongly nonlinear problem, particularly when considering the maximum stress in a strict manner.
We tackle this challenging problem with data-driven topology design.

Figure~\ref{fig_ex4_dd} shows the design domain and boundary conditions of example~4.
Here, we aim to design compliant grippers that transmit force on the input port to an object set on the output port.
In this example, a horizontal load is applied to the input port, and an artificial spring is set on the output port to represent the reaction force from the object.
The magnitude of the applied load per unit area is set to $0.08$ and the spring coefficient of the artificial spring per unit area is set to $10$.
The upper half domain is treated as the actual design domain by imposing a symmetric boundary condition.
In this example, we introduce three objective functions: the material volume, the maximum value of the von Mises stress, and the reaction force on the output port.

The design domain is discretized with $49736$ triangular elements with the representative length of $0.005$.
Concerning the constitutive law, we use the Kirchhoff-Saint Venant material model, and Young's modulus and Poisson's ratio are set to 1 and 0.3, respectively. 
Similar to example~2, we use conforming meshes to structural boundaries to accurately compute the stress.
In addition, we completely deactivate the elements in the void domain because weak materials set in the void domain may cause numerical instabilities in the geometric nonlinear analysis \citep{2020_de_leon_01}.
That is, we extract the structural boundary in the design domain and conduct a geometric nonlinear analysis using only the material domain.

It is worth noting that the VAE may generate material distributions in which an FEM solver cannot find the solution because of the geometric nonlinearity.
Then, we can ignore these material distributions because we focus only on promising material distributions.
This is a great advantage of data-driven topology design when compared with sensitivity-based topology optimization methods, because halting the computation before convergence is fatal in their methods.

As described at the beginning of Section~\ref{section_numerical_examples}, we prepare the initial material distributions by solving a simple compliant mechanism design problem assuming the linear strain because we could not prepare a sufficient number of material distributions that function as compliant mechanisms from the database of the formulation support system used in examples~1 and 2.
In the simple problem, we use the design domain and boundary conditions shown in Fig.~\ref{fig_ex4_dd} and maximize the reaction force on the output port while constraining the material volume.
In addition, we constrain the displacement on the input port to avoid forming extremely thin hinges, instead of strictly constraining the maximum stress.
We implement a simple density-based topology optimization method to solve this problem and obtain $196$ material distributions while changing the magnitude of the input load and the allowable upper limit of the volume.
Then, we evaluate these material distributions using the three objective functions under the geometric nonlinear analysis.
As a result, the material distributions shown in Fig.~\ref{fig_ex4_org_mat} are selected as the initial material distributions according to the optimality of the three objective functions.

We conduct data-driven topology design using these initial material distributions and obtain $400$ material distributions, as shown in Fig.~\ref{fig_ex4_opt_mat}, after 50 iterations.
Their performances are shown in Fig.~\ref{fig_ex4_result}.
The computational time per iteration is about $20$ minutes.
Here, three points should be noted.
First, we constrain the volume, maximum value of the von Mises stress, and reaction force by the respective worst values at iteration~0.
Second, the volume shown in Fig.~\ref{fig_ex4_result} indicates the amount of material in the upper half domain.
Third, the reaction force is shown with a negative sign corresponding to the minimization.

As shown in Fig.~\ref{fig_ex4_result}, the performances of the material distributions are drastically improved after conducting data-driven topology design.
To further investigate the obtained results, we focus on the four material distributions shown in Fig.~\ref{fig_ex4_result}, that is, material distributions~A, B, C, and D.
Material distribution~A is a balanced material distribution at iteration~0, and material distributions~B, C, and D at iteration~50 are completely superior to material distribution~A.
Among them, material distribution~B is specialized to the volume.
Similarly, material distributions~C and D are specialized to the maximum value of the von Mises stress and reaction force, respectively.

Figure~\ref{fig_ex4_detail} shows the contour plots of the von Mises stress with the exact deformations concerning these material distributions.
As shown in this figure, the von Mises stress concentrates in the area circled with the red dotted line in material distribution~A.
On the other hand, in material distributions~B, C, and D, the von Mises stress distribution is well balanced on the hinge part and the displacement support boundary by rounding the area circled with the red dotted lines and adjusting the hinge width.
In particular, material distributions~B and D succeed in approximately evenly distributing the von Mises stress on the three parts.
These results seem to be sufficiently reasonable.

\begin{figure}[!t]
\begin{center}
\includegraphics[scale=1.0]{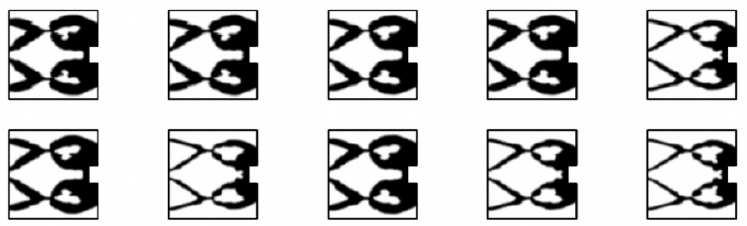}
\caption{Material distributions obtained under linear strain assumption in example~3 (these are a part of the whole)}
\label{fig_ex4_detail_linear}
\end{center}
\end{figure}

If we ignore the geometric nonlinearity by assuming the linear strain, we obtain $400$ high-performance material distributions under this assumption as a result of data-driven topology design.
Figure~\ref{fig_ex4_detail_linear} shows a part of them.
Although these may seem to be reasonable as compliant mechanisms, we cannot conduct a geometric nonlinear analysis of $284$ material distributions, including those in Fig.~\ref{fig_ex4_detail_linear}.
In other words, the linear strain assumption does not hold at least for $71\%$ of the total material distributions.
Furthermore, whereas the maximum reaction force in the remaining $116$ material distributions is $2.92 \times 10^{-3}$, $35$ material distributions achieve a higher reaction force when considering the geometric nonlinearity, as shown in Fig.~\ref{fig_ex4_result}.
Thus, the geometric nonlinearity is not negligible, and a solution search based on a high-fidelity analysis is needed for strongly nonlinear problems, as demonstrated in this example.

At the end of this section, we discuss the relationship between data-driven topology design and \textit{multifidelity topology design} originated by \citet{MP_2020_yaji_01}.
It is a design methodology to indirectly solve a strongly nonlinear topology optimization problem, which is difficult to solve directly and only forward analysis can be conducted, using material distributions obtained by solving another problem, which is easy to solve directly and is correlated with the strongly nonlinear problem.
In their study, the best material distribution among the prepared material distributions was simply selected.
That is, the aim of their study corresponds to obtaining the initial elite solutions in example~3.
In contrast, data-driven topology design can obtain higher-performance elite solutions, as shown in Fig.~\ref{fig_ex4_result}.
Therefore, example~3 indicates that data-driven topology design has potential to be an improved version of multifidelity topology design.

\section{Conclusion}
\label{section_conclusion}
In this paper, we proposed data-driven topology design, which is a sensitivity-free and multi-objective structural design methodology.
It generates higher-performance material distributions from initially provided material distributions using a deep generative model, on the basis of a data-driven manner.
Because of its characteristic, data-driven topology design has potential to enhance a support system for determining an appropriate formulation of a topology optimization problem, and also has potential as a new version of multifidelity topology design.
We demonstrated its usefulness through three numerical examples.
However, some issues remain.

A major one is applying data-driven topology design to various strongly nonlinear problems.
In this paper, we targeted two types of strongly nonlinear problems in structural mechanics as the first investigation of data-driven topology design.
However, there are many other strongly nonlinear design problems.
A representative one is the turbulent flow problem, as described in Section~\ref{section_introduction}.
Therefore, we plan to apply data-driven topology design to the turbulent flow problem.

\begin{figure}[!t]
\begin{center}
\includegraphics[scale=1.0]{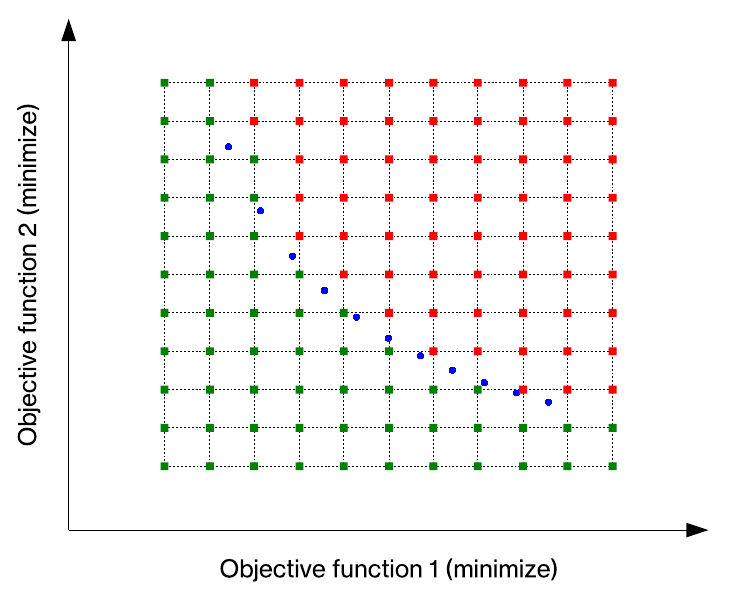}
\caption{An example of fixed grid points for approximately computing the area outside elite solutions colored with blue}
\label{fig_appendix_A}
\end{center}
\end{figure}

\begin{figure}[!t]
\begin{center}
\includegraphics[scale=1.0]{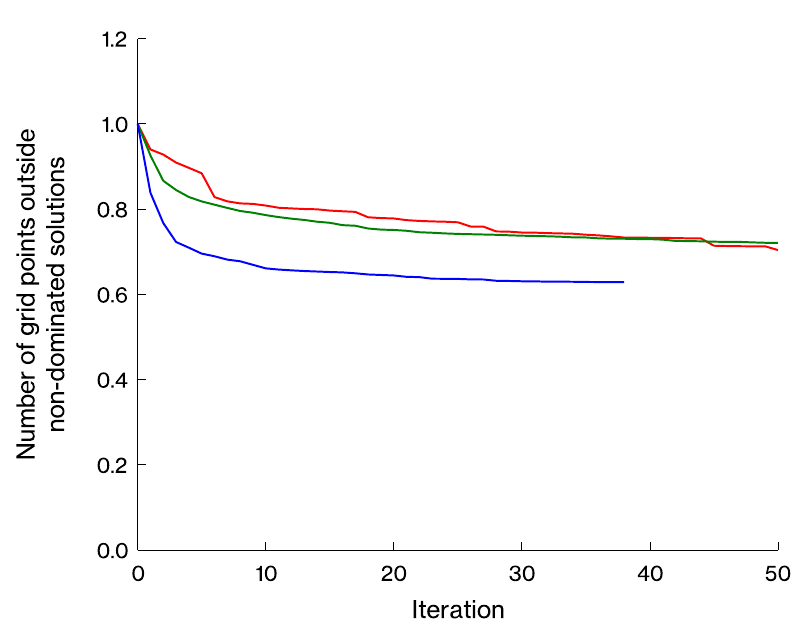}
\caption{Convergence histories of example~1 (blue), example~2 (green), and example~3 (red)}
\label{fig_appendix_A_history}
\end{center}
\end{figure}

As another issue, it is necessary to investigate another deep generative model, despite our adoption of a VAE in this paper; more suitable deep generative models may exist for data-driven topology design.
In addition, a suitable architecture for the VAE should be further investigated.
Although we adopted the architecture shown in Fig.~\ref{fig_VAE_arch} on the basis of the results of a preliminary study, there may be room for improvement.
Furthermore, the theoretical backbone of data-driven topology design should be further investigated.
We plan to tackle these issues in future studies and aim to develop more sophisticated data-driven topology design.

\section{Replication of results}
The necessary information for a replication of the results are presented in the manuscript.
Interested readers may contact the corresponding author for further details regarding the implementation.

\section*{Appendix~A}
The convergence criterion of the proposed implementation is based on the area outside of the elite solutions in the objective function space.
If the area does not decrease after incrementing the iteration, or if the iteration number reaches $50$, we terminate the data generation procedure.

The area outside the elite solutions is approximately computed on the basis of the number of fixed grid points that are not dominated by the elite solutions.
Figure~\ref{fig_appendix_A} shows an example of the fixed grid points for the approximate computation.
In this figure, the red grid points are dominated by the blue elite solutions, and the green grid points are not dominated.
The area outside the blue elite solutions is approximately computed by counting the number of green grid points.

In example~1, the fixed grid points are set by discretizing the objective function space with $400 \times 400$ in the range from $0$ to $1.92$ for the volume and $-1$ to $6$ for the mean compliance.
In example~2, these are set by discretizing the objective function space with $400 \times 400$ in the range from $0$ to $0.7$ for the volume and $0$ to $1.8$ for the von Mises stress.
In example~3, these are set by discretizing the objective function space with $400 \times 400 \times 400$ in the range from $0.1$ to $0.4$ for the volume, $0.1$ to $1.0$ for the von Mises stress, and $-4.0 \times 10^{-3}$ to $-0.5 \times 10^{-3}$ for the reaction force.

Figure~\ref{fig_appendix_A_history} shows the convergence histories of examples~1, 2, and 3.
Note that the number of grid points outside the non-dominated solutions is normalized by the initial values in the respective examples.

\section*{Appendix~B}
\begin{figure}[!t]
\begin{center}
\includegraphics[scale=1.0]{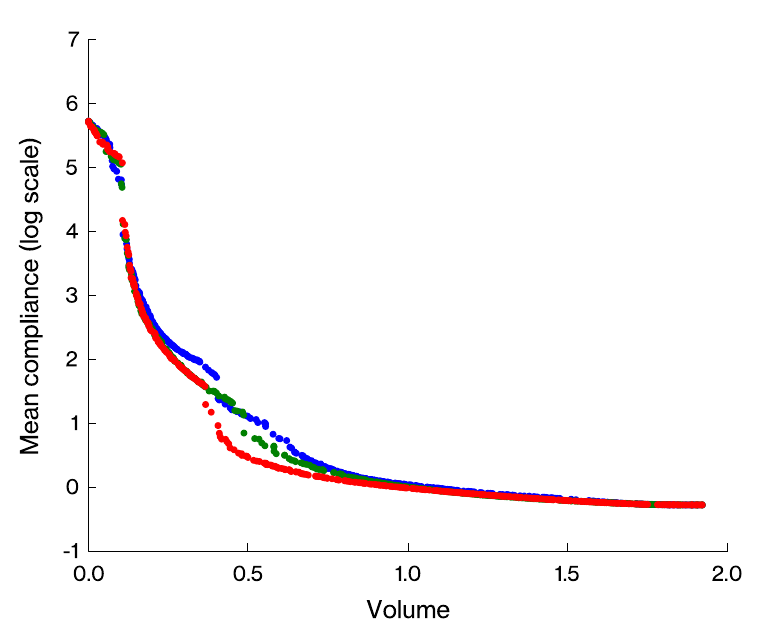}
\caption{Performances of elite solutions obtained while changing the number of the latent variables $N_{\mathrm{lt}}$ in example~1: $N_{\mathrm{lt}} = 2$  (blue), $N_{\mathrm{lt}} = 4$ (green), and $N_{\mathrm{lt}} = 8$ (red)}
\label{fig_appendix_B}
\end{center}
\end{figure}

The number of latent variables $N_{\mathrm{lt}}$ is an important parameter that should be determined carefully.
Here, we investigate its impact on the performances of generated material distributions.
Figure~\ref{fig_appendix_B} shows how $N_{\mathrm{lt}}$ influences obtained results in example~1.
As shown in this figure, the elite solutions are obviously improved by increasing $N_{\mathrm{lt}}$.
Now, we cannot theoretically explain it, but consider that the two- or four-dimensional latent space is too small to capture features of elite material distributions.
On the other hand, if $N_{\mathrm{lt}}$ is set to a larger value, it will be difficult to generate meaningful material distributions by finite random sampling on a large latent space.
Therefore, we assume that there is a suitable range for setting $N_{\mathrm{lt}}$ and adopt $N_{\mathrm{lt}}=8$ for the numerical examples in this paper.

\section*{Conflict of interest}
The authors declare that they have no conflicts of interest.

\bibliographystyle{spbasic}
\bibliography{paper.bbl}

\end{document}